\documentclass[copyright,creativecommons]{eptcs}
\providecommand{\event}{GandALF 2010} 
\usepackage[utf8]{inputenc}
\usepackage{graphicx}
\usepackage{hyperref}
\usepackage{xspace}
\usepackage{color}
\usepackage{amsmath}
\usepackage{amssymb}
\usepackage{stmaryrd}
\usepackage{formal}
\usepackage{tikz}
\usetikzlibrary{automata}
\usetikzlibrary{snakes}
\usetikzlibrary{shapes}
\usetikzlibrary{patterns}
\usetikzlibrary{arrows}
\usepackage{color}
\usepackage{pdfsync}
\usepackage{varioref}
\usepackage{algorithm}
\usepackage{subfigure}
\usepackage{multicol}
\usepackage{amsthm}
\usepackage{rotating}
\usepackage{booktabs,colortbl}
\usepackage{paralist}
\usepackage{courier}
\newcommand{\ie}{i.\,e.~}

\pgfrealjobname{paper}

\begin{document}

\title{Exploiting the Temporal Logic Hierarchy and the Non-Confluence Property for Efficient LTL Synthesis}

\author{Andreas Morgenstern and Klaus Schneider
\institute{%
	University of Kaiserslautern\\
	P.O. Box 3049 \\
	67653 Kaiserslautern, Germany \\
	email: \{morgenstern,schneider\}@cs.uni-kl.de}
}
\def\titlerunning{Efficient $\LTL{}$ Synthesis}
\def\authorrunning{A. Morgenstern \& K. Schneider}

\newtheorem{definition}{Definition}
\newtheorem{proposition}{Proposition}
\newtheorem{lemma}{Lemma}
\newtheorem{theorem}{Theorem}
\newtheorem{corollary}{Corollary}
\newtheorem{remark}{Remark}
\newtheorem{example}{Example}
\def\qed{\hfill \rule{2mm}{2mm}}

\maketitle

\begin{abstract}
The classic approaches to synthesize a reactive system from a linear temporal logic ($\LTL{}$) specification first translate the given $\LTL{}$ formula to an equivalent $\omega$-automaton and then compute a winning strategy for the corresponding $\omega$-regular game. To this end, the obtained $\omega$-automata have to be (pseudo)-determinized where typically a variant of Safra's determinization procedure is used. In this paper, we show that this determinization step can be significantly improved for tool implementations by replacing Safra's determinization by simpler determinization procedures. In particular, we exploit (1) the temporal logic hierarchy that corresponds to the well-known automata hierarchy consisting of safety, liveness, Büchi, and co-Büchi automata as well as their boolean closures, (2) the non-confluence property of $\omega$-automata that result from certain translations of $\LTL{}$ formulas, and (3) symbolic implementations of determinization procedures for the Rabin-Scott and the Miyano-Hayashi breakpoint construction. In particular, we present convincing experimental results that demonstrate the practical applicability of our new synthesis procedure.
\end{abstract}

\begin{section}{Introduction}

In formal verification, we have to check for a given implementation $\cM$ and a given $\LTL{}$ property $\varphi$ whether $\cM$ satisfies $\varphi$ in any environment, which is usually written as $\cM\models\varphi$. In the more general synthesis problem, we have to check whether an (incomplete) implementation\footnote{It is allowed that $\cM$ consists of only the environment.} $\cM$ can be refined by another system (often called a controller) $\cC$ such that a given property $\varphi$ holds, i.e., whether there exists a system $\cC$ such that $\cM\parallel\cC\models\varphi$ holds. This means that the combined behavior of $\cC$ and $\cM$ satisfies $\varphi$ in \emph{any} environment. This controller synthesis problem can be naturally viewed as an infinite game between the controller and an adversary environment.

Due to the enormous progress made during the past two decades, the tools used in formal verification can now be applied to real-world problems, and therefore are routinely used in industrial practice. In contrast, the so-far known tools for the synthesis of $\LTL{}$ specifications can only be applied to small examples which is due to the implementation of these tools: The currently available procedures used in these tools consist of two steps: Similar to verification, the first step consists of translating the $\LTL{}$ formula $\varphi$ to an equivalent nondeterministic $\omega$-automaton $\mA_\varphi$. While this automaton $\mA_\varphi$ can be directly used for symbolic model checking, it is favorable to have a deterministic automaton for constructing a $\omega$-regular game from the obtained automaton $\mA_\varphi$ and the incomplete implementation $\cM$. This determinization step is the main reason for the high complexity of the synthesis problem for $\LTL{}$.

In particular, Safra's construction \cite{Safr88} is often used for this purpose. This construction generates for a given Büchi automaton with $n$ states an equivalent deterministic Rabin automaton with $12^n\cdot n^{2n}$ states and $n$ acceptance pairs. However, Safra's construction is extremely difficult to implement \cite{KlBa06}, and unfortunately not amenable to a symbolic implementation. Indeed, a major drawback of Safra's construction is that known implementations use an \emph{explicit representation} of the automata, since the states of Safra's automaton consist of trees whose nodes are labeled with sets of states. This is probably the main reason why tools for synthesis lack behind (compared to model checking tools that achieved a significant breakthrough first by symbolic state space representations, and second by the use of efficient SAT solvers). As a consequence, tools for $\LTL{}$ synthesis are still limited to very small formulas with only few temporal operators.

In this paper, we show how we can always replace Safra's determinization procedure by simpler ones like the Rabin-Scott subset construction and the Miyano-Hayashi breakpoint construction \cite{MiHa84,Schn03}. To this end, we exploit the membership of the subformulas of a given formula to the classes of the temporal logic hierarchy as well as a recently published determinization procedure for non-confluent $\omega$-automata \cite{MoSc08}.

The \emph{temporal logic hierarchy} \cite{ChMP92,Schn03} has been developed in correspondence to the $\omega$-automata hierarchy \cite{Land69} that has been inspired by the Borel hierarchy known from descriptive set theory. As the $\omega$-automata hierarchy, the temporal logic hierarchy distinguishes between six different classes (cf. Figure~\ref{fig:automata_hierarchy}) of properties. It is moreover known that each class can be characterized by a deterministic class of $\omega$-automata that differ in their acceptance conditions as explained in Section~\ref{AutomataHierarchySec}. While Manna and Pnueli's original definition of the hierarchy was a semantic one (meaning a $\LTL{}$ formula belongs to a class iff there is an equivalent $\omega$-automaton in that class), Schneider \cite{Schn01b,Schn03} presented syntactic characterizations of these classes by means of simple grammars for each of these logics (cf. Figure~\ref{fig:grammar}). Clearly, such syntactic characterizations are incomplete (in the sense that an equivalent formula may belong to a different class), but they are sufficient for practical use: In general, the membership test given by these grammars yields an upper bound of the temporal logic class, and in practice, it often yields the precise class, making therefore more expensive tests unnecessary.

In the following, we will therefore also make use of a syntactic, and therefore incomplete, but more efficiently testable definition to come up with a very efficient $\LTL{}$ synthesis procedure. The key idea of this procedure is thereby based on the following observation: It is well-known since \cite{ChMP92} that for every $\LTL{}$ formula of each temporal logic class, there is an equivalent deterministic automaton with the corresponding acceptance condition. However, it was \emph{previously not recognized} that we can avoid complex determinization procedures like Safra's one once we can express a given $\LTL{}$ formula as a boolean combination of lower class subformulas.

In fact, by the syntactic representation, each formula in the highest class is already a boolean combination of co-Büchi properties, so that we can obtain a Safra-less determinization procedure by first computing deterministic co-Büchi automata for these subformulas, and then computing the boolean closure of the obtained deterministic automata (which is simple as outlined in \cite{Schn03}). The translation of the subformulas to symbolically represented non-deterministic co-Büchi automata has been presented in detail in \cite{Schn01b,Schn03}, and their determinization can be performed using the Miyano-Hayashi breakpoint construction \cite{MiHa84,Schn03}. We can even improve this procedure by determining tighter classes like safety and liveness properties so that even the Rabin-Scott subset construction is sufficient for the determinization.

The second major ingredient of our efficient synthesis procedure are \emph{symbolic determinization constructions} we have developed in \cite{MoSL08} for the Rabin-Scott subset construction \cite{RaSc59} and the Miyano-Hayashi breakpoint construction: For a given symbolically represented nondeterministic automaton, we directly construct symbolic descriptions of the corresponding deterministic automata. Although we can not avoid one exponential step (namely the enumeration of the reachable states of the nondeterministic automaton), we achieved that the symbolic description of the deterministic automaton can be obtained without building it explicitly. Thus, we avoid the enumeration of the exponentially larger state space of the \emph{deterministic} automaton.

Thus, we are able to translate every formula of the temporal logic hierarchy to an equivalent deterministic $\omega$-automaton. Due to results of \cite{ChMP92}, one can translate every $\LTL{}$ formula to a formula contained at least in the highest class of this hierarchy. However, all known such translations require a determinization step, and therefore, they are not useful for our purpose. Thus, we currently have the restriction that the given $\LTL{}$ formulas must already syntactically belong to one of the classes of the hierarchy. In practice, we found that this is almost always the case (in more than 95\% in our benchmarks), and in some other cases, it was not too difficult to rewrite the formula to achieve this membership (checking the equivalence of the rewritten $\LTL{}$ formulas is no problem by verification tools).

To handle the remaining rare cases of formulas that can not be easily rewritten to classes of the temporal logic hierarchy, we use as a final ingredient of our synthesis procedure our recently published determinization procedure for non-confluent automata \cite{MoSc08}. This procedure can be applied to any given $\LTL{}$ formula, provided we use a translation to $\omega$-automata that generates non-confluent automata (which exists!).

{\em In this paper, we build upon the mentioned results we developed in our previous work, i.e., we first try to  decompose a given specification into subformulas that syntactically belong to a class of the temporal logic hierarchy, and apply symbolic implementations of the Rabin-Scott and the Miyano-Hayashi constructions to translate these subformulas to deterministic $\omega$-automata. If this is not directly possible, we make use of a simple translation for full $\LTL{}$ to non-confluent automata so that we can apply our recently published determinization procedure \cite{MoSc08}.}

Having obtained a deterministic automaton, we use standard automata translations to obtain (generalized) deterministic parity automata for each subformula which has to be combined in one conjunctive generalized parity condition. The obtained generalized parity automaton yields a generalized parity game which is solved in the final step of our algorithm using the symbolic algorithm given in \cite{ChHP07}.

The outline of this paper is at follows: after starting with more details on related work (Section~\ref{RelatedWork}) and some basic definitions (Section~\ref{BasicDefinitions}), we explain how we can make use of the temporal logic hierarchy (Section~\ref{TemporalLogicHierarchySec}). The second ingredient is the exploitation of the non-confluence property that is explained in Section~\ref{NonConfluenceProperty}. The added value and the core of this paper is the combination of these results in Section~\ref{SymbolicControllerSynthesis} to obtain an efficient symbolic synthesis procedure for full $\LTL{}$. We conclude the paper by experimental results in Section~\ref{ExperimentalResults}.
\end{section}

\begin{section}{Related Work}

\label{RelatedWork}
There are already symbolic implementations of the subset and breakpoint construction \cite{AEFK05,BCPR06}. In \cite{BCPR06}, procedures are described to compute a symbolically represented nondeterministic automaton from a given alternating automaton, i.e., a non-determinization procedure. Although there are some similarities to our procedure, non-determinization of alternating automata and determinization of nondeterministic automata is different for $\omega$-automata \cite{TuSc05a}. Closer to our determinization procedure is \cite{AEFK05} which generates a deterministic automaton for the safety fragment, and thus implements the subset construction. However, they also start with an alternating automaton which is then translated to an explicitly represented nondeterministic automaton. The nondeterministic automaton is generated on the fly, thus avoiding the construction of the whole explicit automaton. However, this step crucially relies on a translation from alternating automata to the corresponding nondeterministic automata while our procedure is independent of the previous translation from temporal logic to nondeterministic automata. In particular, it is not obvious how the work \cite{AEFK05} could be generalized to more expressive classes like co-Büchi automata.

Since it became clear that the determinization step is the major hurdle in the synthesis of full $\LTL{}$, a recent research trend aims at avoiding determinization by somehow integrating the synthesis procedure with lightweight `pseudo'-determinization procedures. In \cite{KuVa05a}, Kupferman and Vardi present an approach that avoids Safra's determinization and goes through universal co-Büchi word and weak alternating tree automata instead. This approach has been refined in \cite{KuPV06} to allow compositional synthesis or to allow also controller synthesis. A major disadvantage of the approach presented in \cite{KuVa05a} regarding controller synthesis is that it suffers from an exponential blowup with respect to the size of the system under control, even if this system is deterministic. Our approach avoids this exponential blowup. Jobstmann and Bloem developed in \cite{JoBl06} optimizations for this Safra-less approach and developed the tool `Lily'. This tool was the first implementation that is able to synthesize designs that satisfy arbitrary $\LTL{}$ specifications. Although Kupferman and Vardi's approach is potentially amenable to a symbolic implementation, the tool Lily is implemented explicitly, so that is also limited to small $\LTL{}$ formulas only.

In practice, specifications are not given as a single large formula; instead they consist of several relatively small subformulas. In \cite{SoSR08}, an algorithm for $\LTL{}$ synthesis is presented that assumes that the overall specification is given as a conjunction of $\LTL{}$ formulas. Instead of performing determinization for the whole specification, the algorithm generates deterministic automata using the approach of \cite{Pite06} explicitly. Those explicitly represented automata are then encoded symbolically to obtain a generalized parity game which is then solved using the generalized parity algorithm given in \cite{ChHP07a}. Our algorithm assumes a similar setting: We also assume that the specification is a conjunction of $\LTL{}$ formulas. The determinization is also performed only on the nondeterministic automata obtained from these small subformulas, and the final automaton for the overall specification is obtained by combining the single deterministic automata. The main difference is however that we never represent the automata explicitly, so that we expect that our algorithm scales much better (unfortunately, the tool mentioned in \cite{SoSR08} is not publicly available so that we can not perform comparisons).
\end{section}
\begin{section}{Basic Definitions}
\label{BasicDefinitions}
\subsection{Linear Temporal Logic}

For a given set of Boolean variables (propositions) $\Var{\Sigma}$, we define the set of $\LTL{}$ formulas by the following grammar\footnote{We neglect past temporal operators, although these are also available in our framework.}:
$\varphi := \Var{\Sigma}
	\mid \neg \varphi
	\mid \varphi \vee \varphi
	\mid \NEXT{\varphi}
	\mid \SUNTIL{\varphi}{\varphi}
$.
Additionally, we define $\varphi \wedge \psi$, $\EVENTUAL{\varphi}$, $\ALWAYS{\varphi}$, and $\UNTIL{\psi}{\varphi}$ as abbreviations for $\neg (\neg \varphi \vee \neg \psi)$, $\SUNTIL{\varphi}{\True}$, $\neg \EVENTUAL{\neg \varphi}$, and $\SUNTIL{\psi}{\varphi}\vee \ALWAYS{\varphi}$, respectively.

To define the semantics of an $\LTL{}$ formula $\varphi$, we consider infinite sequences of truth assignments to its atomic propositions, i.e., words in $(2^{\Var{\Sigma}})^\omega$. We denote the $(i+1)$-th element of $w$ by $w_i$, i.e., $w=(w_0,w_1,\ldots)$. The semantics of $\LTL{}$ formulas is then defined as follows:
\begin{itemize}
\item for $p \in \Var{\Sigma}$, we have $w,i \models p$ iff $p \in w_i$
\item $w,i \models \neg \varphi$ iff $w,i \not\models \varphi$
\item $w,i \models \varphi \vee \psi$ iff $w,i \models \varphi$ or $w,i\models \psi$
\item $w,i \models \NEXT{\varphi}$ iff $w,i+1 \models \varphi$
\item $w,i \models \SUNTIL{\psi}{\varphi}$ iff there exists $k\geq i$ such that $w,k \models \psi$ and for all $j$ with $i\leq j$ and $j<k$, we have $w,j \models \varphi$
\end{itemize}
For a formula $\varphi$ and a position $j\geq 0$ such that $w,j\models \varphi$ holds, we say that $\varphi$ holds at position $j$ of $w$. If $w,0 \models \varphi$ holds, we say that $\varphi$ holds on $w$, and write simply $w \models \varphi$.

\subsection{$\omega$-Automata on Infinite Words}

\begin{definition}[$\omega$-Automata]
Let $Q$ be a finite set of state variables. Let $\Prop{\Sigma}$ be a finite set of input variables disjoint from $Q$ that defines an alphabet $\Sigma=\Pot{\Prop{\Sigma}}$. Then, a \emph{$\omega$-automaton} $\mA=(\cS,\cI,\cR,\lambda,\cA)$ over the alphabet $\Sigma$ is given by a finite set of states $\cS$, a set of initial states $\cI \subseteq \cS$, a transition relation $\cR \subseteq \cS \times \Sigma \times \cS$, a labeling function $\lambda:\cS \rightarrow \Pot{Q}$ with $\lambda(s) \neq \lambda(s')$ for $s \neq s' \in \cS$ and an acceptance component $\cA$.
\end{definition}

\noindent Using standard terminology, we say that $\mA$ is \emph{deterministic}, if exactly one initial state exists and for each $s \in \cS$ and each input $\sigma \in \Sigma$ there exists exactly one $s' \in \cS$ with  $(s,\sigma,s') \in \cR$.

The acceptance of a word is defined with respect to the set of runs: Given an automaton $\mA=(\cS,\cI,\cR,\cL,\alpha)$ and an infinite word $\alpha:\Nat \rightarrow\Sigma$ over $\Sigma$. Each infinite word $\beta:\Nat \rightarrow \cS$ with $\Let{\beta}{0}\in \cI$ and $\forall t. \left(\Let{\beta}{t},\Leta{t},\Let{\beta}{t+1}\right) \in \cR$ is called a run of $\alpha$ through $\mA$. The set of all runs of $\alpha$ through $\mA$ is defined as follows: We extend the labeling function to runs by defining $\lambda(\beta)=\lambda(\timed{\beta}{0})\lambda(\timed{\beta}{1})\dots$.
Note that $\lambda(\beta)$ is a path over $Q$, called the \emph{trace} of $\beta$. Since every trace of a automaton $\mA$ over a word $\alpha$ is a path over $Q$, we can use $\LTL{}$ to specify the acceptance conditions for $\omega$-automata. To simplify notation, we identify each run with its trace and write $\beta \models \Phi$ to mean that $\lambda(\beta) \models \Phi$. Another form of acceptance conditions are parity conditions that are conveniently defined by a priority function $\Omega:\cS \rightarrow \{0,1,\dots d-1\}$ for some $d \in \Nat$. A run is accepted by a parity condition if the minimal priority seen infinitely often is even. A generalized (conjunctive) parity condition is given by $k$ priority functions and the run is accepted iff it is accepted by each parity condition separately.

Since the states are uniquely identified by the state variables (we have $\lambda(s)\neq \lambda(s')$ for $s \neq s'$)
we can represent each state (set) with a propositional formula.
Hence, we can already represent the set of initial states by a propositional formula $\Phi_\cI$ over $Q \cup \Var{\Sigma}$.
Introducing moreover for each variable $p \in Q $ a next-state variable $p'$ allows us to represent also
the transition relation by a propositional formula $\Phi_\cR$ over $Q\cup\Var{\Sigma}\cup\{p' \mid p \in Q \}$.
\end{section}
\begin{section}{Exploiting the Temporal Logic Hierarchy}
\label{TemporalLogicHierarchySec}

\subsection{The Automata Hierarchy}
\label{AutomataHierarchySec}
In the past, several kinds of acceptance conditions have been proposed and their different expressivenesses have been studied in depth. In particular, the following acceptance conditions have been considered \cite{Wagn79,Thom90a,Schn03}:

\begin{itemize}
\item A run is accepted by a safety condition $\ALWAYS{\varphi}$ with a state
set $\varphi$ if the run exclusively runs through the set $\varphi$.
\item A run is accepted by a liveness condition $\EVENTUAL{\varphi}$ with
a state set $\varphi$ if the run visits at least one state of the set $\varphi$ at least once.
\item A run is accepted by a Büchi condition $\ALWAYS{\EVENTUAL{\varphi}}$ with
a state set $\varphi$ if the run visits at least one state of the set $\varphi$
infinitely often.
\item A run is accepted by a co-Büchi condition $\EVENTUAL{\ALWAYS{\varphi}}$
with a state set $\varphi$ if the run visits only states of the set $\varphi$
infinitely often.
\end{itemize}

\noindent The above conditions define the corresponding automaton classes ${\sf (N)Det}_{\ALWAYS{}}$, ${\sf (N)Det}_{\EVENTUAL{}}$, ${\sf (N)Det}_{\ALWAYS{\EVENTUAL{}}}$, and ${\sf (N)Det}_{\EVENTUAL{\ALWAYS{}}}$, respectively. Moreover, ${\sf (N)Det}_{\sf Prefix}$ and ${\sf (N)Det}_{\sf Streett}$ automata have acceptance conditions of the form $\bigwedge_{j=0}^{f} \ALWAYS{\varphi_j} \vee \EVENTUAL{\psi_j}$ and $\bigwedge_{j=0}^{f}\ALWAYS\EVENTUAL{\varphi_j}\vee \EVENTUAL\ALWAYS{\psi_j}$, respectively.

The expressiveness of these classes is illustrated in Figure~\ref{fig:automata_hierarchy}, where $\lseqexpress{\cC_1}{\cC_2}$ means that for any automaton in $\cC_1$, there is an equivalent one in $\cC_2$. Moreover, we define $\eqexpress{\cC_1}{\cC_2} := \lseqexpress{\cC_1}{\cC_2} \wedge \lseqexpress{\cC_2}{\cC_1}$ and $\lsexpress{\cC_1}{\cC_2} := \lseqexpress{\cC_1}{\cC_2} \wedge \neg (\eqexpress{\cC_1}{\cC_2})$. As can be seen, the hierarchy consists of six different classes, and each class has a deterministic representative.

\begin{figure}
\tikzstyle{tikblock} = [draw, fill=blue!6, rounded corners]
\begin{tikzpicture}[shorten >=1pt,node distance=2cm,auto]
    	 \node[tikblock,minimum height=2cm] (G) at (0,5) {$\begin{array}{c}
	 							\text{\sf NDet}_{\sf \ALWAYS{}}\\
	 							\text{\sf Det}_{\sf \ALWAYS{}}
									\\\vspace{\baselineskip}
									\\\vspace{\baselineskip}
								\end{array}$};
	 \node[tikblock,minimum height=2cm] (F) at (0,1)  {$\begin{array}{c}
	 							\text{\sf NDet}_{{\sf \EVENTUAL{}}}^\text{\sf total}\\
								\text{\sf Det}_{{\sf \EVENTUAL{}}}
									\\\vspace{1\baselineskip}
								\end{array}$};
	\node[tikblock,minimum height=2cm] (Prefix) at (4,3) {$\begin{array}{c}
								\text{\sf Det}_\text{Prefix}
									\\\vspace{1.5\baselineskip}
								\end{array}$};
	\node[tikblock,minimum height=2cm] (GF) at (8,5) {$\begin{array}{c}
	 							\text{\sf Det}_{{\ALWAYS\EVENTUAL{}}}
									\\\vspace{\baselineskip}
									\\\vspace{\baselineskip}
								\end{array}$};
	\node[tikblock,minimum height=2cm] (FG) at (8,1) {$\begin{array}{c}
	 							\text{\sf NDet}_\text{Prefix}\\
	 							\text{\sf NDet}_{\EVENTUAL{}}\\
	 							\text{\sf (N)Det}_{\EVENTUAL{\ALWAYS{}}}
							    \\\vspace{\baselineskip}
								\\\vspace{\baselineskip}
								\end{array}$};
	\node[tikblock,minimum height=2cm] (Streett) at (12,3) {$\begin{array}{c}
	 							\text{\sf NDet}_{{\ALWAYS\EVENTUAL{}}} \\
								\text{\sf (N)Det}_{\text{Streett}} \\
								\text{\sf (N)Det}_{\text{Parity}}
									\\\vspace{1\baselineskip}
									\\\vspace{1\baselineskip}\\
								\end{array}$};
								
	\node[tikblock,fill=blue!20,minimum width=1.5cm] at (0.025,4.5) {$\text{\sf TL}_{\ALWAYS{}}$};
	\node[tikblock,fill=blue!20,minimum width=1.5cm] at (0.025,0.4) {$\text{\sf TL}_{\EVENTUAL{}}$};
	\node[tikblock,fill=blue!20,minimum width=1.5cm] at (4,2.4) {$\text{\sf TL}_\text{\sf Prefix}$};
	\node[tikblock,fill=blue!20,minimum width=1.5cm] at (12,2.3) {$\text{\sf TL}_{\text{Streett}}$};
	\node[tikblock,fill=blue!20,minimum width=1.4cm] at (8,4.5) {$\text{\sf TL}_{\ALWAYS{\EVENTUAL{}}}$};
	\node[tikblock,fill=blue!20,minimum width=1.5cm] at (8,0.4) {$\text{\sf TL}_{\EVENTUAL{\ALWAYS{}}}$};
	
	\path[->,color=white] (G) edge node [sloped,left=-0.3cm] {\color{black} $\lsexpress{}{}$} (Prefix)
		     (F) edge node [sloped,right=-0.3cm] {\color{black} $\lsexpress{}{}$} (Prefix)
		     (Prefix) edge node [sloped,right=-0.4cm] {\color{black} $\lsexpress{}{}$} (GF)
		     (Prefix) edge node [sloped,left=-0.4cm] {\color{black}  $\lsexpress{}{}$} (FG)	
		     (GF) edge node [sloped,left=-0.4cm] {\color{black} $\lsexpress{}{}$} (Streett)
		     (FG) edge node [sloped,right=-0.3cm] {\color{black} $\lsexpress{}{}$} (Streett)	;		
\end{tikzpicture}	
\caption{(Borel) Hierarchy of $\omega$-Automata and Temporal Logic}
\label{fig:automata_hierarchy}
\end{figure}
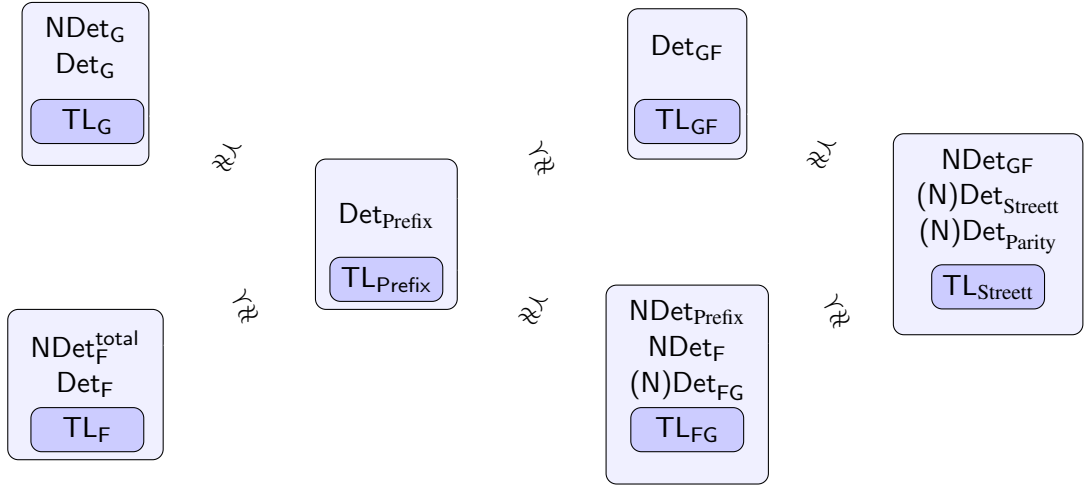

\subsection{The Temporal Logic Hierarchy}

In \cite{ChMP92,Schn01b,Schn03}, corresponding hierarchies for temporal logics have been defined.  Following \cite{Schn01b,Schn03}, we define the hierarchy of temporal logic formulas syntactically by the grammar rules of Fig.~\ref{fig:grammar}:

\begin{figure*}[!th] \[
\begin{array}{|c|c|}
\hline
	\begin{array}{ll}
	P_{\ALWAYS{}} ::= &
		\Var{\Sigma} \\ &
		\mid \neg P_{\EVENTUAL{}}
		\mid P_{\ALWAYS{}}\wedge P_{\ALWAYS{}}
		\mid P_{\ALWAYS{}}\vee P_{\ALWAYS{}} \\ &
		\mid \NEXT{P_{\ALWAYS{}}}
		\mid \ALWAYS{P_{\ALWAYS{}}} \\ &
		\mid \UNTIL{P_{\ALWAYS{}}}{P_{\ALWAYS{}}} \\
	\end{array} &
	\begin{array}{ll}
	P_{\EVENTUAL{}} ::= &
		\Var{\Sigma} \\ &
		\mid \neg P_{\ALWAYS{}}
		\mid P_{\EVENTUAL{}}\wedge P_{\EVENTUAL{}}
		\mid P_{\EVENTUAL{}}\vee P_{\EVENTUAL{}} \\ &
		\mid \NEXT{P_{\EVENTUAL{}}}
		\mid \EVENTUAL{P_{\EVENTUAL{}}} \\ &
		\mid \SUNTIL{P_{\EVENTUAL{}}}{P_{\EVENTUAL{}}} \\
	\end{array}
\\
\hline \multicolumn{2}{|c|}{%
	\begin{array}{ll}
	P_{\sf Prefix} ::= &
		P_{\ALWAYS{}}
		\mid P_{\EVENTUAL{}}
		\mid \neg P_{\sf Prefix}
		\mid P_{\sf Prefix} \wedge P_{\sf Prefix}
		\mid P_{\sf Prefix} \vee P_{\sf Prefix}
	\end{array}}
\\\hline
	\begin{array}{ll}
	P_{\ALWAYS{\EVENTUAL{}}} ::= &
		P_{\sf Prefix} \\ &
		\mid \neg P_{\EVENTUAL{\ALWAYS{}}}
		\mid P_{\ALWAYS{\EVENTUAL{}}}\wedge P_{\ALWAYS{\EVENTUAL{}}}
		\mid P_{\ALWAYS{\EVENTUAL{}}}\vee P_{\ALWAYS{\EVENTUAL{}}} \\ &
		\mid \NEXT{P_{\ALWAYS{\EVENTUAL{}}}}
		\mid \ALWAYS{P_{\ALWAYS{\EVENTUAL{}}}} \\ &
		\mid \UNTIL{P_{\ALWAYS{\EVENTUAL{}}}}{P_{\ALWAYS{\EVENTUAL{}}}}
		\mid \SUNTIL{P_{\EVENTUAL{}}}{P_{\ALWAYS{\EVENTUAL{}}}}
	\end{array} &
	\begin{array}{ll}
	P_{\EVENTUAL{\ALWAYS{}}} ::= &
		P_{\sf Prefix} \\ &
		\mid \neg P_{\ALWAYS{\EVENTUAL{}}}
		\mid P_{\EVENTUAL{\ALWAYS{}}}\wedge P_{\EVENTUAL{\ALWAYS{}}}
		\mid P_{\EVENTUAL{\ALWAYS{}}}\vee P_{\EVENTUAL{\ALWAYS{}}} \\ &
		\mid \NEXT{P_{\EVENTUAL{\ALWAYS{}}}}
		\mid \EVENTUAL{P_{\EVENTUAL{\ALWAYS{}}}} \\ &
		\mid \SUNTIL{P_{\EVENTUAL{\ALWAYS{}}}}{P_{\EVENTUAL{\ALWAYS{}}}}
		\mid \UNTIL{P_{\EVENTUAL{\ALWAYS{}}}}{P_{\ALWAYS{}}}
	\end{array}
\\
\hline \multicolumn{2}{|c|}{%
	\begin{array}{ll}
	P_{\sf Streett} ::= &
		P_{\ALWAYS{\EVENTUAL{}}}
		\mid P_{\EVENTUAL{\ALWAYS{}}}
		\mid \neg P_{\sf Streett}
		\mid P_{\sf Streett} \wedge P_{\sf Streett}
		\mid P_{\sf Streett} \vee P_{\sf Streett}
\end{array}
} \\
\hline
\end{array}
\]
\caption{Syntactic Characterizations of the Classes of the Temporal Logic Hierarchy}
\label{fig:grammar}
\end{figure*}

\begin{definition}[Temporal Logic Classes] \label{temp_borel_1_def}
For $\kappa\in\{\ALWAYS{}$, $\EVENTUAL{}$, ${\sf Prefix}$, $\EVENTUAL{\ALWAYS{}}$, $\ALWAYS{\EVENTUAL{}}$, ${\sf Streett} \}$, we define the logics ${\sf TL}_\kappa$ by the grammars given in Fig.~\ref{fig:grammar}, where ${\sf TL}_\kappa$ is the set of formulas that can be derived from the nonterminal $P_\kappa$ ($\Var{\Sigma}$ represents any variable $v\in\Var{\Sigma}$).
\end{definition}

\noindent Typical safety conditions like $\ALWAYS{\varphi}$ or $\ALWAYS{\UNTIL{b}{a}}$ that state that something bad never happens, are contained in ${\sf TL}_\ALWAYS{}$. Liveness conditions like $\EVENTUAL{\varphi}$ are contained in ${\sf TL}_\EVENTUAL{}$. Finally, fairness conditions like $\ALWAYS\EVENTUAL{\varphi}$ that demand that something good infinitely often happens, are contained in ${\sf TL}_{\ALWAYS\EVENTUAL{}}$ while stabilization/persistence properties like $\EVENTUAL\ALWAYS{\varphi}$ that demand that after a finite interval, nothing bad happens are contained in ${\sf TL}_{\EVENTUAL\ALWAYS{}}$.

In our experience, almost all formulas that occur in practice belong to ${\sf TL}_{\sf Streett}$. If a given formula should not belong to ${\sf TL}_{\sf Streett}$, it is often straightforward to rewrite it to an equivalent ${\sf TL}_{\sf Streett}$ formula. For example, the formula $\ALWAYS\EVENTUAL{\UNTIL{a}{b}}$ that demands that infinitely often $\UNTIL{a}{b}$ holds is equivalent to the ${\sf TL}_{\sf Streett}$ formula $\ALWAYS\EVENTUAL{\SUNTIL{a}{b}} \vee \EVENTUAL\ALWAYS{\UNTIL{a}{b}}$. Clearly, there are many formulas outside ${\sf TL}_{\sf Streett}$, but we claim that these formulas seldom occur in practice.

\subsection{Relating the Temporal Logic and the Automata Hierarchy}

In \cite{Schn01b,Schn03} several translation procedures are given to translate formulas from ${\sf TL}_{\kappa}$ to equivalent ${\sf (N)Det}_\kappa$ automata. In particular, the following is an important result:

\begin{theorem}[Temporal Logic and Automaton Hierarchy] \label{borel_thm}
Given a formula $\Phi \in {\sf TL}_\kappa$, we can construct a deterministic $\omega$-automaton $\mA=(\Pot{Q},\cI,\cR,\lambda,\cA)$ of the class ${\sf Det}_\kappa$ in time $O(2^{\card{\Phi}})$ with $\card{Q}\leq 2^{\card{\Phi}}$ state variables. Therefore, $\mA=(\Pot{Q},\cI,\cR,\lambda,\cA)$ is a symbolic representation of a deterministic automaton with $O(2^{2^{\card{\Phi}}})$ states.
\end{theorem}

\noindent The above results are already proved in detail in \cite{Schn03}, where translation procedures from ${\sf TL}_\kappa$ to ${\sf NDet}_\kappa$ have been constructed. Moreover, it has been shown in \cite{Schn03} that the subset construction can be used to determinize the automata that stem from the classes ${\sf TL}_{\ALWAYS{}}$ and ${\sf TL}_{\EVENTUAL{}}$ and that the Miyano-Hayashi breakpoint construction is sufficient to determinize the automata that stem from the translation of formulas from ${\sf TL}_{\EVENTUAL{\ALWAYS{}}}$ and ${\sf TL}_{\ALWAYS{\EVENTUAL{}}}$.

Since ${\sf TL}_{\sf Prefix}$ and ${\sf TL}_{\sf Streett}$ are the boolean closures of ${\sf TL}_{\ALWAYS{}}\cup{\sf TL}_{\EVENTUAL{}}$ and ${\sf TL}_{\EVENTUAL{\ALWAYS{}}}\cup{\sf TL}_{\ALWAYS{\EVENTUAL{}}}$, respectively, the remaining results for ${\sf TL}_{\sf Prefix}$ and ${\sf TL}_{\sf Streett}$ follow from the boolean combinations of ${\sf Det}_{\ALWAYS{}}/{\sf Det}_{\EVENTUAL{}}$ and ${\sf Det}_{\EVENTUAL{\ALWAYS{}}}/{\sf Det}_{\ALWAYS{\EVENTUAL{}}}$, respectively.

The final step consists of computing the boolean closure of the acceptance conditions. To this end, it is shown in \cite{Schn03} how arbitrary boolean combinations of $\ALWAYS{\varphi}$ and $\EVENTUAL{\varphi}$ with propositional formulas $\varphi$ are translated to equivalent ${\sf Det}_{\sf Prefix}$ automata, and analogously, how arbitrary boolean combinations of $\ALWAYS{\EVENTUAL{\varphi}}$ and $\EVENTUAL{\ALWAYS{\varphi}}$ with propositional formulas $\varphi$ are translated to equivalent ${\sf Det}_{\sf Streett}$ automata.

\end{section}

\begin{section}{Exploiting the Non-Confluence Property}
\label{NonConfluenceProperty}

It is well-known that the $\omega$-automata that stem from $\LTL{}$ formulas are a special class that has already found several characterizations. Due to results of \cite{McPa71}, the automata can be characterized as \emph{non-counting} automata, and in terms of alternating automata, the class of \emph{linear weak} or \emph{very weak} automata has been defined \cite{MeSe03,MuSS86b}. Moreover, many translation procedures from $\LTL{}$ generate {\em unambiguous automata} \cite{CaMi03} where every accepted word has a unique accepting run \cite{Schn03} (although there may be additional non-accepting runs for the same word). The determinization procedure presented in this chapter makes use of the fact that the automata generated from $\LTL{}$ are unambiguous. Without useless states, the transition relation of an unambiguous automaton has a certain form that we call \emph{non-confluence}:

\begin{center}
\begin{minipage}{13cm}
\emph{An automaton is non-confluent if whenever two runs of the same infinite word meet at a state $q$, then they must share the entire finite prefix up to state $q$.}
\end{minipage}
\end{center}

\noindent To give an intuitive idea why the automata constructed from $\LTL{}$ formulas by the `standard translation' procedure are non-confluent, consider the automaton of Figure~\ref{fig:LTLtranslate1} that is obtained by translating the formula $\SUNTIL{\psi}{\varphi}$ to a non-deterministic automaton.
As explained in \cite{Schn01b,Schn03}, the `standard' translation procedure from LTL to $\omega$-automata traverses the syntax tree of the LTL formula in a bottom-up manner and abbreviates each subformula that starts with a temporal operator. The subformula $\SUNTIL{\psi}{\varphi}$ is thereby abbreviated by a new state variable $q$, and the preliminary transition relation $\cR$ is replaced with $\cR\wedge \left(q \leftrightarrow \psi\vee\varphi\wedge\BNEXT{q}\right)$.

\begin{figure}
\centering
\begin{tikzpicture}[node distance=3cm,auto]
\node[state] (empty){$\overline{q}$};
\node[state,right of=empty] (q) {$q$};
\path[->]
      	(empty) edge  [bend left] node        {$\overline{\varphi} \wedge\overline{\psi}$} (q)
      	(empty) edge [loop left] node {$\overline{\psi}$}  ()
	(q) edge [loop right] node {$\varphi \vee \psi$}  ()
	(q) edge  [bend left] node        {$\psi$} (empty)
	;
	
 \node [right=1 cm,text width=8cm]  at (q)
   {
   	\[
    	\begin{array}{|cc||l|}
		\hline
		\psi & \varphi & q \leftrightarrow \psi\vee\varphi\wedge\BNEXT{q} \\
		\hline\hline
		\False 	& \False	& q \leftrightarrow \False\\
		\False 	& \True		& q \leftrightarrow \BNEXT{q}\\
		\True 	& \False	& q \leftrightarrow \True\\
		\True 	& \True		& q \leftrightarrow \True\\
		\hline
	\end{array}
	\]
   };
\end{tikzpicture}
\caption{Nondeterministic Automaton for $\SUNTIL{\psi}{\varphi}$}
\label{fig:LTLtranslate1}
\end{figure}

As can be seen by Figure~\ref{fig:LTLtranslate1}, the input $\varphi\wedge\neg\psi$ demands that the current state is maintained, but allows the automaton to be in any of the two states. The other three classes of inputs uniquely determine the current state, but leave the successor state completely unspecified\footnote{According to the Krohn-Rhodes decomposition theorem, automata that stem from $\LTL{}$ properties do only have reset and identity inputs \cite{Schn03}.}. As a consequence, input words that infinitely often satisfy $\neg(\varphi\wedge\neg\psi)$, i.e., $\neg\varphi\vee\psi$, do only have one (infinite) run, while the remaining input words that satisfy $\varphi\wedge\neg\psi$ from a certain point of time on do have two infinite runs that are of the form $\xi{q}^\omega$ and $\xi\overline{q}^\omega$ with the same finite prefix $\xi$. Hence, the automaton is non-confluent, since the
two runs never merge after they have split.

An example run tree (that encodes all the runs of a given word) is shown in Fig.~\ref{LTL_Runtree1_fig}. It can be seen that there is a uniquely determined run, since all other nondeterministic choices lead to finite paths. Another example run tree that contains two infinite runs is shown in Figure~\ref{LTL_Runtree2_fig}.

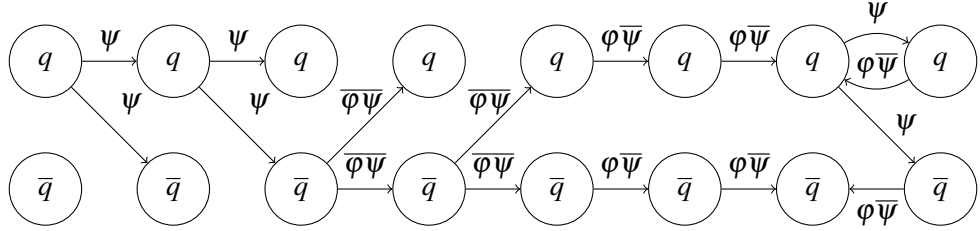
\begin{figure}
\centering
\begin{tikzpicture}[node distance=1.7cm,auto]
\node[state] (q0) {$q$};
\node[state,below of=q0] (q0') {$\overline{q}$};
\node[state,right of=q0] (q1) {$q$};
\node[state,below of=q1] (q1') {$\overline{q}$};
\node[state,right of=q1] (q2) {$q$};
\node[state,below of=q2] (q2') {$\overline{q}$};
\node[state,right of=q2] (q3) {$q$};
\node[state,below of=q3] (q3') {$\overline{q}$};
\node[state,right of=q3] (q4) {$q$};
\node[state,below of=q4] (q4') {$\overline{q}$};
\node[state,right of=q4] (q5) {$q$};
\node[state,below of=q5] (q5') {$\overline{q}$};
\node[state,right of=q5] (q6) {$q$};
\node[state,below of=q6] (q6') {$\overline{q}$};
\node[state,right of=q6] (q7) {$q$};
\node[state,below of=q7] (q7') {$\overline{q}$};
\path[->]
	(q0) edge node {$\psi$} (q1)
	(q0) edge node {$\psi$} (q1')
	(q1) edge node {$\psi$} (q2)
	(q1) edge node {$\psi$} (q2')
	(q2') edge node [left=0.5mm,above=0mm]{$\overline{\varphi}\overline{\psi}$} (q3)
	(q2') edge node {$\overline{\varphi}\overline{\psi}$} (q3')
	(q3') edge node [left=0.5mm,above=0mm]{$\overline{\varphi}\overline{\psi}$} (q4)
	(q3') edge node {$\overline{\varphi}\overline{\psi}$} (q4')
	(q4) edge node {$\varphi\overline{\psi}$} (q5)
	(q4') edge node {$\varphi\overline{\psi}$} (q5')
	(q5) edge node {$\varphi\overline{\psi}$} (q6)
	(q5') edge node {$\varphi\overline{\psi}$} (q6')
	(q6) edge [bend left] node {$\psi$} (q7)
	(q7) edge [bend left] node [above=0mm]{$\varphi\overline{\psi}$} (q6)
	(q6) edge node [above=0mm,right=1mm] {$\psi$} (q7')
	(q7') edge node {$\varphi\overline{\psi}$} (q6')
	
	;

\end{tikzpicture}
\caption{Run Tree with a Uniquely Determined Infinite Run}
\label{LTL_Runtree1_fig}
\end{figure}

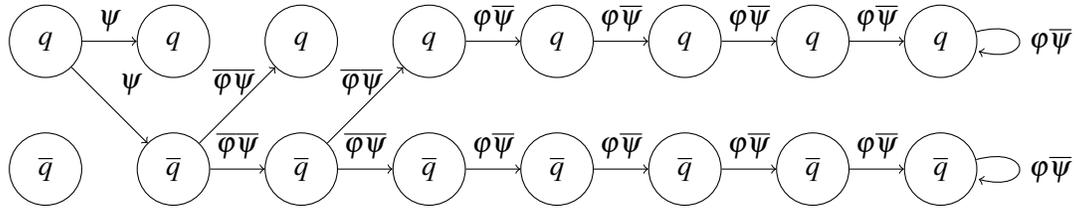
\begin{figure}
\centering
\begin{tikzpicture}[node distance=1.7cm,auto]
\node[state] (q0) {$q$};
\node[state,below of=q0] (q0') {$\overline{q}$};
\node[state,right of=q0] (q1) {$q$};
\node[state,below of=q1] (q1') {$\overline{q}$};
\node[state,right of=q1] (q2) {$q$};
\node[state,below of=q2] (q2') {$\overline{q}$};
\node[state,right of=q2] (q3) {$q$};
\node[state,below of=q3] (q3') {$\overline{q}$};
\node[state,right of=q3] (q4) {$q$};
\node[state,below of=q4] (q4') {$\overline{q}$};
\node[state,right of=q4] (q5) {$q$};
\node[state,below of=q5] (q5') {$\overline{q}$};
\node[state,right of=q5] (q6) {$q$};
\node[state,below of=q6] (q6') {$\overline{q}$};
\node[state,right of=q6] (q7) {$q$};
\node[state,below of=q7] (q7') {$\overline{q}$};
\path[->]
	(q0) edge node {$\psi$} (q1)
	(q0) edge node {$\psi$} (q1')
	(q1') edge node [left=0.5mm,above=0mm]{$\overline{\varphi}\overline{\psi}$} (q2)
	(q1') edge node {$\overline{\varphi}\overline{\psi}$} (q2')
	(q2') edge node [left=0.5mm,above=0mm]{$\overline{\varphi}\overline{\psi}$} (q3)
	(q2') edge node {$\overline{\varphi}\overline{\psi}$} (q3')
	(q3) edge node {$\varphi\overline{\psi}$} (q4)
	(q3') edge node {$\varphi\overline{\psi}$} (q4')
	(q4) edge node {$\varphi\overline{\psi}$} (q5)
	(q4') edge node {$\varphi\overline{\psi}$} (q5')
	(q5) edge node {$\varphi\overline{\psi}$} (q6)
	(q5') edge node {$\varphi\overline{\psi}$} (q6')
	(q6) edge node {$\varphi\overline{\psi}$} (q7)
	(q6') edge node {$\varphi\overline{\psi}$} (q7')
	(q7) edge [loop right] node {$\varphi\overline{\psi}$}()
	(q7') edge [loop right] node {$\varphi\overline{\psi}$} ()
	
	;

\end{tikzpicture}
\caption{Run Tree with two Infinite Runs of the Automaton of Fig.~\ref{fig:LTLtranslate1}}
\label{LTL_Runtree2_fig}
\end{figure}

As every automaton $\mA_\varphi$ obtained by the translation of a $\LTL{}$ formula $\varphi$ is a product of non-confluent automata, and as the product automaton of two non-confluent automata is also non-confluent, it follows that the automata $\mA_\varphi$ obtained by the above `standard' translation are non-confluent.

The above non-confluence property has been used in \cite{MoSc08} to develop a determinization procedure that exploits symbolic set representations. In particular, it does not rely on Safra trees as used by Safra's original procedure \cite{Safr88} or by the improved version of Piterman \cite{Pite06}. The states of the deterministic automata obtained by these procedures are trees of subsets of states of the original automaton. In contrast, our procedure generates deterministic automata whose states consist of $n$-tuples of subsets of states, where $n$ is the number of states of the nondeterministic automaton.

\end{section}
\begin{section}{A Symbolic Controller Synthesis Algorithm for Full $\LTL{}$}
\label{SymbolicControllerSynthesis}

\subsection{The Averest Framework}

Averest \cite{Schn09} is a set of tools for the specification, verification, and implementation of reactive systems. It includes a compiler and a simulator for synchronous programs, a symbolic model checker, and a tool for hardware-software synthesis. Averest can be used for modeling and verifying finite as well as infinite state systems at various levels of abstraction. In particular, Averest is not only well-suited for hardware design, but also for modeling communication protocols, concurrent programs, software in embedded systems, etc.

The design flow using Averest consists of the following steps: First, the system is described as a synchronous program in our language Quartz \cite{Schn09}, a descendent of Esterel. Then, the program is translated to a transition system in the Averest Interchange Format (AIF) using the Quartz compiler. This intermediate description can be directly used for verification with the symbolic model checker to check whether the system satisfies its specifications. If this is the case, code can be generated for an implementation in hardware or software with output formats like VHDL, Verilog or C.

The compiler contained in Averest does not only compile a Quartz program to a corresponding transition system, but also provides procedures to translate $\LTL{}$ and other specification logics to symbolically represented $\omega$-automata.

The tool implementing the features of this paper is called Opal and extends the current Averest compiler to deal with controller synthesis. In the following subsections, the different steps of Opal are described in more detail. Opal is implemented in Moscow ML and uses Moscow MLs foreign function interface to connect to the BDD-package CUDD.

\subsection{Decomposing Specifications}

Specifications often consist of several relatively simple components - for instance, a collection of $\LTL{}$ properties whose conjunction should be satisfied. Thus, we consider specifications of the form $\Phi = \bigwedge_{j=0}^N \Phi_j$.

Instead of translating the entire specification at once, we generate separate deterministic automata for every subformula $\Phi_j$. Clearly, since we allow any $\LTL{}$ property, we may have to make use of the determinization procedure outlined in \cite{MoSc08} to translate $\Phi_j$. For example, this is the case if the top-level operator of $\Phi_j$ is a temporal operator and $\Phi_j$ does not belong to one of the lower Borel classes ${\sf TL_{\ALWAYS{\EVENTUAL{}}}}$ or ${\sf TL_{\EVENTUAL{\ALWAYS{}}}}$.

In practice, this is however nearly never the case. Instead, most often, also the subformulas $\Phi_j$ are a boolean combination of even smaller formulas. Instead of handling them all at once, we break also these into smaller parts so that we obtain for every $\Phi_j$ a collection $\varphi_1\dots \varphi_k$ of $\LTL{}$ properties that all start with a temporal operator, and that either belong to one of the classes ${\sf TL_{\ALWAYS{}}}$, ${\sf TL_{\EVENTUAL{}}}$, ${\sf TL_{\ALWAYS{\EVENTUAL{}}}}$, ${\sf TL_{\EVENTUAL{\ALWAYS{}}}}$ or to none of these classes.

\subsection{Handling Safety and Liveness formulas}

Safety formulas $\varphi_i$ are first translated to a nondeterministic safety automaton. Although safety automata can not be directly transformed to a parity automaton, it is possible to minimize them using direct simulation\footnote{Indeed, the nondeterministic automata generated by the translation procedures given in \cite{Schn03} translates the safety fragment to an automaton with the trivial acceptance condition $\ALWAYS{\true}$, so that even the ordinary simulation relation as described in \cite{EtWS05} could be applied.}. After minimization, we perform the ordinary subset construction and afterwards minimize the automaton again using the direct simulation relation. However, one important subclass of properties does not scale well using this approach. Since many specifications are of the form `if something now happens, in the next step something else happens', we treat this subclass separately. This subclass can be formally described by Boolean combinations of formula of the form $\psi$ or of the form $\NEXT{\xi}$ where both $\psi$ and $\xi$ are Boolean formulas over the input variables.

In that case, every $\NEXT{}$ operator doubles the state space of the non-deterministic automaton and thus leads to a blow-up in the number of BDD variables of the deterministic automaton. Even worse, the simulation relations can neither minimize the nondeterministic nor the deterministic automaton since the two states that occur due to an occurrence of $\NEXT{a}$ can not be equivalent as one of the two will lead to a non-satisfying loop. Thus, the basic translation procedure really suffers from a double-exponential blowup. Instead, we translate those formulas by abbreviating each variable $a$ that is not under the scope of a $\NEXT$ operator by a previous variable $a_{p}$ such that $\BNEXT{a_p} \leftrightarrow a$ holds and replace any subformula $\NEXT{a}$ with $a$.

\begin{proposition}
Given a formula $\ALWAYS{\varphi}$ where $\varphi$ is a boolean combination of state variables $v \in \Var{\Sigma}$ and formulas $\NEXT{v}$ where $v \in \Var{\Sigma}$. Then $\ALWAYS{\varphi}$ is initially equivalent to the symbolically represented deterministic automaton $\mA=(\Pot{Q},\Phi_\cI,\Phi_\cR,\cL,\cA)$ where $Q=\{a_p \mid a \in \Prop{}\} \cup \{p\}$, $\Phi_\cI=p \wedge \bigwedge_{a \in \Var{\Sigma}} \neg a_p$, $\cA=\ALWAYS(p)$ and the transition relation is defined by
\[\Phi_\cR=
  \begin{array}{l}
  \displaystyle{\bigwedge_{a \in \Var{\Sigma}} \BNEXT{a_p} \leftrightarrow a \wedge \BNEXT{p}\leftrightarrow\varphi'}
  \end{array}
\]
Here $\varphi'$ is obtained from $\varphi$ by replacing any occurrence of $\BNEXT{a}$ with $a$ and any occurrence of a formula $a \in \Prop{}$ with $a_p$. Moreover $\cL: 2^Q \rightarrow 2^Q$ is the identity function. \end{proposition}

Obviously, this leads to an automaton that has at most $\card{a}+1$ state variables, thus the automaton is only exponential in the size of the specification. Nevertheless, if many subformulas $a$ exist but only little subformulas $\NEXT{a}$, the ordinary translation may give better results, so that this translation is optional in our algorithm.

For liveness formulas $\varphi_i$, we translate $\neg \varphi_i$ and dualize the corresponding deterministic safety automaton to obtain a deterministic liveness automaton.

\subsection{Handling Co-Büchi and Büchi specifications}

For co-Büchi specifications, we use the translation from ${\sf TL_{\EVENTUAL{\ALWAYS{}}}}$ to nondeterministic co-Büchi automata, minimize this automaton using the minimization techniques of \cite{Frit05b}. Afterwards this automaton is determinized using the breakpoint construction and again minimized. Büchi specifications are translated using the dual deterministic automaton of the formula obtained from negating the formula.

\subsection{Handling $\LTL{}$ Formulas that do not belong to a lower Borel Class}

If a subformula $\Phi_i$ does not belong to one of the lower Borel classes, we have to resort on the determinization procedure from \cite{MoSc08}. Thus, we first translate $\Phi_i$ to a non-confluent Büchi automaton and minimize it. This nondeterministic automaton is determinized using the procedure described in \cite{MoSc08}. However, we do not need to construct the whole automaton at once. Instead, we construct the automaton for a fixed bound $k$ and check whether every marking of a state has been noticed by a state set. If so, we return this automaton, otherwise we do the same with an increased bound. Afterwards, we use the minimization techniques of \cite{Frit05b} to minimize the obtained parity automaton.

Although this procedure gives back a parity automaton that is from a theoretical point of view more efficient than a Street automaton, the heavy complexity of determinization makes even this approach infeasible in practice. Thus, we break up also the formulas $\Phi_i$ into smaller parts until every subformula starts with a temporal operator. Those subformulas are then translated as explained before. It is well known that every parity automaton can also be interpreted as a Streett or a Rabin automaton. We thus interpret the obtained automaton as a Streett automaton and combine the deterministic Streett automata to obtain a Streett automaton for $\Phi_i$ that is afterwards translated to a generalized parity automaton.

\subsection{Solving Generalized Parity Games}
Instead of solving the whole generalized parity game at once, we first solve the subgames that are obtained by constructing the game for the subformula $\Phi_j$. The set of states that are loosing for the controller need not be considered in the overall game. Afterwards, we solve the reduced overall game using the generalized parity game algorithm of \cite{ChHP07a}.

\subsection{Generating Circuits from BDDs}
The output of the generalized parity algorithm is a BDD over the (current state) variables $\VarU,\VarC,\StateG$ and over newly introduced state variables $\Var{M}$ to encode counter variables that are used to switch between the sub-strategies calculated by the generalized parity algorithm. A slight modification of the algorithm given in Figures 2 and 3 of \cite{BGJP07a} allows us to generate for every controllable input variable $c$ a BDD $\varphi_c$ with the meaning that $c$ should hold whenever $\varphi_c$ holds. We then write those BDDs to a file in our Averest interchange format. The tool Topaz in our Averest toolset can be used to obtain either Verilog, VHDL or C code from the generated file.

\end{section}
\begin{section}{Experiments and Conclusion}
\label{ExperimentalResults}
This section describes the experiments performed using the controller synthesis algorithm described in this paper. To this end, the $23$ specifications that come with the Lily tool \cite{JoBl06} are used as a benchmark set. Those 23 handwritten formulas are mostly traffic light examples or arbiters.

To analyze the effect of using the all-purpose determinization of \cite{MoSc08} instead of the much simpler breakpoint or subset construction, we performed three different experiments on each of the formulas where the runtimes are summarized in Figure~\ref{table:det_table}. The first column lists the identification number of the example. The second column gives some measures of the specification in terms of number of temporal operators, boolean operators, and the number of input variables. The third column gives the number of subformulas, \ie the number of top-level conjunctions. The conjunction of those formula form the overall specification. Notice that the number of boolean operators do not include the top-level conjuncts. The next column indicates how many subformulas are not safety formulas\footnote{In the considered experiments, any subformula is at least contained in the Büchi or co-Büchi class so that we obtain a formula of the temporal logic hierarchy.}. The next two columns give the time for determinization followed by the overall time needed to synthesize the specification.

Algorithm Opal uses all optimizations. Every formula is translated with the lowest possible determinization procedure. In contrast, algorithm Nonc translates every formula that is not a safety formula with the all-purpose determinization construction of \cite{MoSc08}. Finally, the last column gives the running time of the Lily algorithm.

\begin{figure}
\centering
\beginpgfgraphicnamed{experiment}
	\pgfkeys{/pgf/number format/.cd,fixed,precision=2}
	\pgfplotstableset{columns={No.,{T,B,AP},subf,O,{NoncDetT},{OpalDetT},{NoncT},{OpalT},{LilyT}}}
	\pgfplotstabletypeset
	[
		every even row/.style={ before row={\rowcolor[gray]{0.9}}},
		every head row/.style={before row=\toprule,after row=\midrule},
		every last row/.style={after row=\bottomrule},
		col sep=semicolon,
		columns/No./.style={string type},
		columns/{T,B,AP}/.style={string type},
		columns/{p}/.style={string type},
	]{Experiments.table}
\endpgfgraphicnamed
\caption{Experiments performed with different Determinization Constructions}
\label{table:det_table}
\end{figure}

As it can be seen, using "'easier"' determinization procedures like the Rabin-Scott subset construction or the Miyano-Hayashi breakpoint construction significantly improves the determinization step itself. When the size of the specification grows as in examples L20-L23, the overall synthesis time is no longer dominated by the determinization, but instead by the time to solve the generalized parity game. In that case using the easier determinization procedures leads to smaller sized automata and thus also to an improvement on the overall synthesis time. With the exception of the small specifications L1 and L2, our tool is significantly faster compared to the explicitly implemented tool Lily.

\end{section}
\bibliographystyle{eptcs}
\bibliography{paper}
\end{document}